%% file: 15_07_03_FD.tex
\documentclass[conference,romanappendices,10pt]{IEEEtran}	

\IEEEoverridecommandlockouts

\usepackage{algorithm,algorithmic,amsmath,amssymb,amsthm,array,bibentry,cite,color,comment,enumerate}
\usepackage{enumitem,eurosym,float,graphicx,lettrine,mathrsfs,multicol,multirow,nomencl,pict2e,psfrag}
\usepackage{ragged2e,setspace,stfloats,subfigure,supertabular,tabularx,url}
\usepackage[USenglish]{babel}

{		\newtheorem{theorem}{Theorem}}
{ 		}
{ 		\newtheorem{lemma}{Lemma}}
{ 		}
{ 		\newtheorem{remark}{Remark}}
{		}
{		}
{ 		\newtheorem{corollary}{Corollary}}
{		

\input{./Definitions.tex}

\begin{document}

\title{Full-Duplex MIMO Small-Cell Networks: \\ Performance Analysis}

		
\author{\IEEEauthorblockN{Italo Atzeni and Marios Kountouris}%
		\IEEEauthorblockA{Mathematical and Algorithmic Sciences Lab, France Research Center, Huawei Technologies Co. Ltd. \\
		\{italo.atzeni, marios.kountouris\}@huawei.com}}

\maketitle

\begin{abstract}
Full-duplex small-cell relays with multiple antennas constitute a core element of the envisioned 5G network architecture. In this paper, we use stochastic geometry to analyze the performance of wireless networks with full-duplex multi-antenna small cells, with particular emphasis on the probability of successful transmission. To achieve this goal, we additionally characterize the distribution of the self-interference power of the full-duplex nodes. The proposed framework reveals useful insights on the benefits of full-duplex with respect to half-duplex in terms of network throughput.
\end{abstract}

\begin{IEEEkeywords}
Full-duplex, multiple antennas, small cells, performance analysis, stochastic geometry. 
\end{IEEEkeywords}

\section{Introduction} \label{sec:Intro}

The employment of full-duplex (FD) technology in wireless networks has been recognized as a promising solution to cope with the ever-growing demand for high data rates.
A key drawback of FD wireless communication is the self-interference received at the FD nodes from their own transmission. However, thanks to recent advances in self-interference mitigation techniques \cite{Sab14}, it is now possible to implement FD radios in practical settings \cite{Bha13}. In this regard, small-cell (SC) systems prove especially suitable for the deployment of FD technology due their low transmit powers and the low mobility of their users \cite{Ngu14}.

Several recent works, such as \cite{Ton15,Lee15}, have examined the performance of hybrid FD/HD large-scale networks, despite considering only single-antenna nodes; in addition, the self-interference channel gain has been modeled as a constant value, which is a very coarse approximation and is only meaningful when digital cancellation is applied \cite{Ton15,Rii11}. On the one hand, it is timely and relevant to investigate FD nodes with multiple antennas in view of the promising concept of FD multiple-input multiple-output (MIMO) relays \cite{Rii11}, even if the extension from the single-antenna to the multiple-antenna case sensibly complicates the analysis. On the other hand, the residual self-interference channel is known to be subject to Ricean fading\footnote{Before applying active cancellation, the magnitude of the self-interference channel can be modeled as a Ricean distribution with large $K$-factor due to the strong line-of-sight component; after applying active cancellation, the line-of-sight component is reduced, resulting in smaller $K$-factor \cite{Dua12}.} and, therefore, its modeling in a MIMO context represents a challenging problem when receive combining and transmit beamforming techniques are employed. In addition, the precise knowledge of the distribution of the self-interference power is essential for a rigorous system-level performance analysis.

In this paper, we fill these gaps by providing the following contributions: i) using powerful tools from stochastic geometry, we study the performance of wireless networks with randomly distributed FD MIMO nodes and derive tight bounds for the probability of successful transmission; ii) we characterize the distribution of the FD self-interference power under Ricean fading for arbitrary receive and transmit beamforming strategies. In our setting, the FD nodes resemble small-cell relays, which are envisioned to be at the foundation of 5G \cite{Jun14}. Finally, numerical results are reported~to corroborate our theoretical findings and to establish under which conditions the employment of FD is beneficial for the network.



\begin{figure}[t!]
\centering
\includegraphics[scale=1]{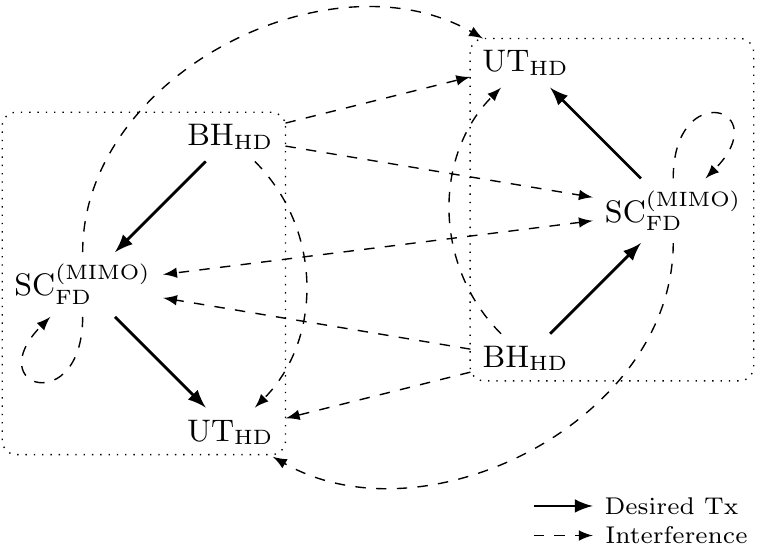} 
\caption{System model with full-duplex MIMO small cells (SC$_{\mathrm{FD}}^{(\mathrm{MIMO})}$), half-duplex backhaul base stations (BH$_{\mathrm{HD}}$), and half-duplex user terminals (UT$_{\mathrm{HD}}$), with corresponding desired transmissions and interferences.} \label{fig:topology} \vspace{-3mm}
\end{figure}

\section{System Model} \label{sec:SM}
\subsection{System Setup} \label{sec:SS}

Consider the scenario in Figure~\ref{fig:topology}, where a set of FD SCs acts as relays between a set of backhaul (BH) base stations (BSs) and a set of mobile user terminals (UTs), both operating in half-duplex (HD) mode; all communications occur in the same frequency band. In our setting, the SCs are equipped with multiple antennas, whereas the UTs have a single antenna; on the other hand, a multiple antenna BH BS performing space division multiple access (SDMA) and sending one stream to the SC can be considered, hence being equivalently seen as single-antenna BS by each SC. During a given time-slot, we assume that each SC communicates exactly with one desired BH BS and with one desired UT. The analysis of the downlink (i.e., from the BH BS to the mobile UT) and of the uplink (i.e., from the mobile UT to the BH BS) are equivalent since they both consist of a succession of a single-input multiple-output and a multiple-input single-output transmissions: therefore, in the following, we generalize our model and refer to HD transmitting and receiving nodes.

Let us thus introduce the stationary, independently marked Poisson point process (PPP) $\Phi_{m} \triangleq \big\{ (x_{n}, \widetilde{m}(x_{n}), \widehat{m}(x_{n})) \big\}$ on $\Real^{2} \times \Real^{2} \times \Real^{2}$. We use $\Phi \triangleq \{ x_{n} \}$ to denote the PPP of the FD nodes with spatial density $\lambda$; likewise, $\widetilde{\Phi} \triangleq \widetilde{m}(\Phi) = \{ \widetilde{m}(x) \}_{x \in \Phi}$ and $\widehat{\Phi} \triangleq \widehat{m}(\Phi) = \{ \widehat{m}(x) \}_{x \in \Phi}$ are the isotropic marks of $\Phi$ denoting the HD transmitting and receiving nodes, respectively, with fixed distances of the desired links given by $\| x - \widetilde{m}(x) \| = \widetilde{R}$ and $\| x - \widehat{m}(x) \| = \widehat{R}$, $\forall x \in \Phi$: evidently, $\widetilde{\Phi}$ and $\widehat{\Phi}$ are PPPs dependent on $\Phi$ and have also density $\lambda$. In our scenario, it is reasonable to assume that $\widetilde{R} \gg \widehat{R}$, since the SCs cover a rather small area compared to the range of the BH BSs. For convenience, in the rest of the paper we use the notation $\widetilde{m}_{x} \triangleq \widetilde{m}(x)$ and $\widehat{m}_{x} \triangleq \widehat{m}(x)$.

\subsection{Channel Model} \label{sec:CM}

We assume that the FD nodes and the HD transmitting nodes transmit with constant power $\rho$ and $\widetilde{\rho}$, respectively; furthermore, the FD nodes are equipped with $N_{R}$ receive antennas and $N_{T}$ transmit antennas. The propagation through the wireless channel is characterized as the combination of a pathloss attenuation and a small-scale fading. The pathloss function between the nodes $x$ and $z$ given by $l(z,x) \triangleq \|z - x\|^{-\alpha}$, with pathloss exponent $\alpha > 2$.

Given transmitting node $x$ and receiving node $z$, we use the following notation.\footnote{To improve readability, in the rest of the paper, $x$ always refers to a transmitting node and $z$ to a receiving node (either FD or HD).} The channels are denoted as $\H_{x z} \in \Compl^{N_{R} \times N_{T}}$ if $x, z \in \Phi$, as $\h_{x z} \in \Compl^{N_{R}}$ if $x \in \widetilde{\Phi}$ and $z \in \Phi$, as $\h_{x z} \in \Compl^{N_{T}}$ if $x \in \Phi$ and $z \in \widehat{\Phi}$, and as $h_{x z} \in \Compl$ if $x \in \widetilde{\Phi}$ and $z \in \widehat{\Phi}$; in particular, $\H_{x x}$ models the self-interference at $x \in \Phi$ resulting from its own transmission. Furthermore, $s_{x}$ represents the data symbol transmitted by $x$ with $\Exp \{ |s_{x}|^2 \} = 1$, whereas the additive noise is denoted by $\n_{z} \in \Compl^{N_{R}}$ if $z \in \Phi$ and by $n_{z} \in \Compl$ if $z \in \widehat{\Phi}$, with elements distributed independently as $\setC \setN (0, \sigma^{2})$. Lastly, $\v_{z} \in \Compl^{N_{R}}$ indicates the receive combining vector applied by $z \in \Phi$ and $\w_{x} \in \Compl^{N_{T}}$ is the transmit beamforming vector applied by $x \in \Phi$, with $\| \v_{z} \|^{2} = \| \w_{x} \|^{2} = 1$.

We assume that all the channels, except the self-interference channel, are subject to Rayleigh fading with elements distributed independently as $\setC \setN (0, 1)$. On the other hand, the self-interference channel is subject to Ricean fading \cite{Dua12} and, therefore, the elements of $\H_{x x}$ are distributed independently as $\setC \setN (\mu, \nu^{2})$. In this regard, one can measure the Ricean $K$-factor and the self-interference attenuation $\Omega$ and determine the mean and standard deviation of $\H_{x x}$ as (cf. \cite{Tep03})
\begin{align} \label{eq:mu_nu}
\mu \triangleq \sqrt{\frac{K \Omega}{K+1}}, \qquad \nu \triangleq \sqrt{\frac{\Omega}{K+1}}.
\end{align}

\subsection{SINR Characterization} \label{sec:SINR}

In this section, we characterize the signal-to-interference-plus-noise ratio (SINR) for the FD nodes and the HD receiving nodes, which are needed in the next section to analyze the probability of successful transmission. In doing so, we study the first hop (i.e., the SINR at the FD nodes) and the second hop (i.e., the SINR at the HD receiving nodes) separately under the assumption that the whole point process is reshuffled after the first hop (see details in Section~\ref{sec:SP}).

\noindent \textbf{First Hop:} Consider a reference FD node indexed by $k$. Building on Slivnyak's theorem \cite[Ch.~8.5]{Hae12}, we assume that this is located at the origin and, due to the stationarity of $\Phi$, the statistics of its signal reception are seen by any FD node: we can thus write $l(k,x) = |X_{x}|^{-\alpha}$, with $|X_{x}|$ being the distance of $x$ from the origin. Hence, the received signal at FD node $k$ (its desired transmitter being $\widetilde{m}_{k}$) is given by
\begin{align}
\nonumber \y_{k} \triangleq & \ \underbrace{\sqrt{\widetilde{\rho}} \widetilde{R}^{-\frac{\alpha}{2}} \h_{\widetilde{m}_{k} k} s_{\widetilde{m}_{k}}}_{\mathrm{(a)}} + \sum_{\ell \in \Phi \backslash \{ k \}} \underbrace{\sqrt{\rho} |X_{\ell}|^{-\frac{\alpha}{2}} \H_{\ell k} \w_{\ell} s_{\ell}}_{\mathrm{(b)}} \\
\label{eq:y_1} & \hspace{-6mm} + \sum_{\ell \in \Phi \backslash \{ k \}} \underbrace{\sqrt{\widetilde{\rho}} |X_{\widetilde{m}_{\ell}}|^{-\frac{\alpha}{2}} \h_{\widetilde{m}_{\ell} k} s_{\widetilde{m}_{\ell}}}_{\mathrm{(c)}} + \underbrace{\sqrt{\rho} \H_{k k} \w_{k} s_{k}}_{\mathrm{(d)}} + \n_{k}
\end{align}
where (a) represents the desired signal, (b) and (c) indicate the interference coming from FD node $\ell$ and its associated HD transmitting node $\widetilde{m}_{\ell}$, respectively, and (d) represents the self-interference. Given the receive combining vector $\v_{k}$, the resulting SINR reads as
\begin{align} \label{eq:SINR_1}
\mathrm{SINR}_{k} & \triangleq \frac{\widetilde{\rho} \widetilde{R}^{-\alpha} S_{\widetilde{m}_{k} k}}{I_{k} + \sigma^{2}}
\end{align}
where we have defined
\begin{equation}
S_{x k} \triangleq \left\{
\begin{array}{ll}
|\v_{k}^{\herm} \H_{x k} \w_{x}|^{2}, & x \in \Phi \\
|\v_{k}^{\herm} \h_{x k}|^{2}, & x \in \widetilde{\Phi}
\end{array} \right.
\end{equation}
and where $I_{k}$ is the overall interference at $k$, i.e.,
\begin{align} \label{eq:I_1}
I_{k} & \triangleq \sum_{\ell \in \Phi \backslash \{k\}} \big( \rho |X_{\ell}|^{-\alpha} S_{\ell k} + \widetilde{\rho} |X_{\widetilde{m}_{\ell}}|^{-\alpha} S_{\widetilde{m}_{\ell} k} \big) + \rho S_{k k}.
\end{align}
The success probability of the first hop is derived in Section~\ref{sec:SP1}.

\noindent \textbf{Second Hop:} Consider a reference HD receiving node indexed by $\widehat{m}_{k}$. Again, following Slivnyak's theorem, we assume that this is located at the origin and, due to the stationarity of $\widehat{\Phi}$, the statistics of its signal reception are seen by any HD receiving node: we can thus write $l(\widehat{m}_{k},x) = |X_{x}|^{-\alpha}$. Hence, the received signal at HD receiving node $\widehat{m}_{k}$ (its desired transmitter being $k$) is given by
\begin{align}
\nonumber y_{\widehat{m}_{k}} \triangleq & \ \underbrace{\sqrt{\rho} \widehat{R}^{-\frac{\alpha}{2}} \h_{k \widehat{m}_{k}}^{\herm} \w_{k} s_{k}}_{\mathrm{(a)}} + \sum_{\ell \in \Phi \backslash \{k\}} \underbrace{\sqrt{\rho} |X_{\ell}|^{-\frac{\alpha}{2}} \h_{\ell \widehat{m}_{k}}^{\herm} \w_{\ell} s_{\ell}}_{\mathrm{(b)}} \\
\label{eq:y_2} & \hspace{16mm} + \sum_{\ell \in \Phi} \underbrace{\sqrt{\widetilde{\rho}} |X_{\widetilde{m}_{\ell}}|^{-\frac{\alpha}{2}} h_{\widetilde{m}_{\ell} \widehat{m}_{k}} s_{\widetilde{m}_{\ell}}}_{\mathrm{(c)}} + n_{\widehat{m}_{k}}
\end{align}
where (a) represents the desired signal, (b) indicates the interference coming from FD node $\ell$, and (c) is the interference from HD transmitting node $\widetilde{m}_{\ell}$. The resulting SINR is given by \vspace{-4mm}
\begin{align} \label{eq:SINR_2}
\mathrm{SINR}_{\widehat{m}_{k}} & \triangleq \frac{\rho \widehat{R}^{-\alpha} S_{k \widehat{m}_{k}}}{I_{\widehat{m}_{k}} + \sigma^{2}}
\end{align}
where we have defined
\begin{equation}
S_{x \widehat{m}_{k}} \triangleq \left\{
\begin{array}{ll}
|\h_{x \widehat{m}_{k}}^{\herm} \w_{x}|^{2}, & x \in \Phi \\
|h_{x \widehat{m}_{k}}|^{2}, & x \in \widetilde{\Phi}
\end{array} \right.
\end{equation}
and where $I_{\widehat{m}_{k}}$ is the overall interference at $\widehat{m}_{k}$, i.e.,
\begin{align} \label{eq:I_2}
I_{\widehat{m}_{k}} \triangleq \sum_{\ell \in \Phi \backslash \{k\}} \rho |X_{\ell}|^{-\alpha} S_{\ell \widehat{m}_{k}} + \sum_{\ell \in \Phi} \widetilde{\rho} |X_{\widetilde{m}_{\ell}}|^{-\alpha} S_{\widetilde{m}_{\ell} \widehat{m}_{k}}.
\end{align}
The success probability of the second hop is derived in Section~\ref{sec:SP2}.

\section{Success Probability} \label{sec:SP}

The successful transmission of a packet over the complete path, i.e., from the HD transmitting node to the HD receiving node through the FD node, is given by the joint ccdf of $\mathrm{SINR}_{k}$ and $\mathrm{SINR}_{\widehat{m}_{k}}$, which is denoted by $\mathrm{P}_{\mathrm{suc}} \triangleq \Pr (\mathrm{SINR}_{k} > \theta, \mathrm{SINR}_{\widehat{m}_{k}} > \theta)$ \cite{Vaz11} for a given SINR threshold $\theta$ (without loss of generality, we consider the same SINR threshold for the two hops). Assuming no correlation between the two hops due to independent sampling of a point process being reshuffled after each hop, i.e., the transmission in the two hops occurs over two uncorrelated instances of $\Phi_{m}$, it follows that
\begin{align} \label{eq:P_suc}
\mathrm{P}_{\mathrm{suc}} = \mathrm{P}_{\mathrm{suc}}^{(1)} \mathrm{P}_{\mathrm{suc}}^{(2)}
\end{align}
where we have defined $\mathrm{P}_{\mathrm{suc}}^{(1)} \triangleq \Pr (\mathrm{SINR}_{k} > \theta)$ and $\mathrm{P}_{\mathrm{suc}}^{(2)} \triangleq \Pr (\mathrm{SINR}_{\widehat{m}_{k}} > \theta)$. The case where the interference is spatio-temporally correlated will be analyzed in a longer version of this paper. Note that, based on the Fortuin-Kasteleyn-Ginibre inequality, the uncorrelated case considered here can be shown to be a lower bound on the network performance. In addition, for notational simplicity, we assume that the system performance is interference-limited and, hence, we consider $I_{k} \gg \sigma^{2}$ and $I_{\widehat{m}_{k}} \gg \sigma^{2}$.

\subsection{First Hop} \label{sec:SP1}

In this section, we analyze the success probability of the first hop $\mathrm{P}_{\mathrm{suc}}^{(1)} = \Pr (\mathrm{SINR}_{k} > \theta)$, i.e., the probability of successful transmission from HD transmitting node $\widetilde{m}_{k}$ to FD node $k$. First, considering $\mathrm{SINR}_{k}$ in \eqref{eq:SINR_1}, we have $S_{\widetilde{m}_{k} k} \sim \chi_{2 N_{R}}^{2}$ (desired signal) and $S_{x k} \sim \chi_{2}^{2}$, $\forall x \in (\Phi \backslash \{ k \}) \cup (\widetilde{\Phi} \backslash \{ \widetilde{m}_{k} \})$ (interferers).\footnote{We define a $\chi_{2 n}^{2}$ random variable to have pdf $f(x) = \frac{x^{n-1} e^{-x}}{\Gamma(n)}$.} The following lemma gives a tight approximation of the distribution of the self-interference power $S_{k k}$.

\begin{lemma} \label{lem:SI} \rm{
The self-interference power $S_{k k} = |\v_{k}^{\herm} \H_{k k} \w_{k}|^{2}$ is Gamma distributed\footnote{We define a Gamma random variable with shape parameter $a$ and scale parameter $b$ to have pdf $f(x) = \frac{x^{a-1} e^{-x/b}}{b^a \Gamma(a)}$.} with shape parameter $a$ and scale parameter $b$ given by
\begin{align} \label{eq:ab}
a \triangleq \frac{(\mu^{2} + \nu^{2})^{2}}{\gamma \mu^{4} + 2 \mu^{2} \nu^{2} + \nu^{4}}, \qquad
b \triangleq \frac{\gamma \mu^{4} + 2 \mu^{2} \nu^{2} + \nu^{4}}{\mu^{2} + \nu^{2}}
\end{align}
respectively, where $\mu$ and $\nu$ are the mean and standard deviation, respectively, of the self-interference channel $\H_{k k}$ (see \eqref{eq:mu_nu}) and where we have defined
\begin{align}
\gamma \triangleq \frac{4 N_{R} N_{T} - (N_{R}+1) (N_{T}+1)}{(N_{R}+1) (N_{T}+1)}.
\end{align}}
\end{lemma}

\begin{IEEEproof}
See Appendix~\ref{sec:A_SI}.
\end{IEEEproof}

\noindent Lemma~\ref{lem:SI} represents a key result of this paper since it provides a formal characterization of the self-interference power experienced by a FD MIMO node with arbitrary beamforming vectors, based on the knowledge of the parameters $K$ and $\Omega$ (whose values are available either by design or by measurements).
					      
The next theorem provides the success probability of the first hop.

\begin{theorem} \label{th:P_suc1} \rm{
The success probability of the first hop is given by \vspace{-4mm}
\begin{align} \label{eq:P_suc1}
\mathrm{P}_{\mathrm{suc}}^{(1)} & \triangleq \sum_{n=0}^{N_{R} - 1} \bigg[ \frac{(- s)^{n}}{n!} \frac{\mathrm{d}^{n}}{\mathrm{d} s^{n}} \mathcal{L}_{I_{k}}(s) \bigg]_{s = \theta \widetilde{\rho}^{-1} \widetilde{R}^{\alpha}}
\end{align}
where \vspace{-2mm}
\begin{align} \label{eq:LI_1a}
\mathcal{L}_{I_{k}}(s) \triangleq \frac{1}{(1 + s b \rho)^{a}} \exp \big( - \lambda \Upsilon(s) \big)
\end{align}
is the Laplace transform of $I_{k}$ (cf. \eqref{eq:I_1}), where we have defined
\begin{align} \label{eq:Upsilon}
\Upsilon (s) \triangleq \int_{0}^{\infty} \bigg( 2 \pi - \frac{1}{1 + s \rho r^{-\alpha}} \Psi(s, r) \bigg) r \mathrm{d} r
\end{align}
with
\begin{align} \label{eq:Psi}
\Psi (s, r) \triangleq \int_{0}^{2 \pi} \frac{\mathrm{d} \varphi}{1 + s \widetilde{\rho} (\widetilde{R}^{2} + r^{2} + 2 \widetilde{R} r \cos \varphi)^{- \frac{\alpha}{2}}}.
\end{align}
}
\end{theorem}

\begin{IEEEproof}
See Appendix~\ref{sec:A_P_suc1_th}.
\end{IEEEproof}

\vspace{1mm}

Given the integral form of $\Upsilon (s)$ in \eqref{eq:Upsilon}, the success probability $\mathrm{P}_{\mathrm{suc}}^{(1)}$ is not in closed-form and needs to be evaluated numerically; nonetheless, we derive the following lower and upper bounds.

\begin{corollary} \label{cor:P_suc1} \rm{
The Laplace transform of $I_{k}$ in \eqref{eq:LI_1a} is bounded as $\mathcal{L}_{I_{k}}(s) \in \big[ \mathcal{L}_{I_{k}}^{(\min)}(s), \mathcal{L}_{I_{k}}^{(\max)}(s) \big]$, with
\begin{align}
\label{eq:LI_1min} \mathcal{L}_{I_{k}}^{(\min)}(s) & \triangleq \frac{1}{(1 + s b \rho)^{a}} \exp \big( - \lambda \Upsilon^{(\max)}(s) \big), \\
\label{eq:LI_1max} \mathcal{L}_{I_{k}}^{(\max)}(s) & \triangleq \frac{1}{(1 + s b \rho)^{a}} \exp \big( - \lambda \Upsilon^{(\min)}(s) \big)
\end{align}
where we have defined
\begin{align}
\label{eq:Upsilon_min} \Upsilon^{(\min)}(s) & \triangleq (1 + \tfrac{2}{\alpha}) (\rho + \widetilde{\rho}) \frac{\pi^{2} s^{\frac{2}{\alpha}}}{\alpha \sin \big( \frac{2 \pi}{\alpha} \big)}, \\
\label{eq:Upsilon_max} \Upsilon^{(\max)}(s) & \triangleq 2 (\rho + \widetilde{\rho}) \frac{\pi^{2} s^{\frac{2}{\alpha}}}{\alpha \sin \big( \frac{2 \pi}{\alpha} \big)}.
\end{align}
Then, the lower and upper bounds of the success probability of the first hop $\mathrm{P}_{\mathrm{suc}}^{(1)}$ are obtained by substituting $\mathcal{L}_{I_{k}}(s)$ in \eqref{eq:P_suc2} with $\mathcal{L}_{I_{k}}^{(\min)}(s)$ and $\mathcal{L}_{I_{k}}^{(\max)}(s)$, respectively.
}
\end{corollary}

\begin{IEEEproof}
See Appendix~\ref{sec:A_P_suc1_cor}.
\end{IEEEproof}

\begin{remark} \rm{
In order to efficiently compute the derivatives of the bounds \eqref{eq:LI_1min}--\eqref{eq:LI_1max}, one can resort to the well-known general Leibniz rule \cite[Eq.~3.3.8]{Abr72} for the differentiation of the product of two functions $f(s) g(s)$: for instance, for $\mathcal{L}_{I_{k}}^{(\min)}(s)$, we can write $f(s) = \frac{1}{(1 + s b \rho)^{a}}$ and $g(s) = \exp \big( -\lambda \Upsilon^{(\max)}(s) \big)$. In turn, the derivatives of $g(s)$ can be computed using Fa\`{a} di Bruno's formula \cite{Joh02} for the differentiation of the composition of two functions $(g_{1} \circ g_{2})(s)$, with $g_{1}(s) = \exp(s)$ and $g_{2}(s) = - \lambda \Upsilon^{(\max)}(s)$. These considerations apply equivalently to the bounds provided in Corollary~\ref{cor:P_suc2}.
}
\end{remark}

The following corollary provides a sufficient condition under which FD outperforms HD in terms of throughput in the case of single-antenna nodes.

\begin{corollary} \label{cor:HD} \rm{
Assume that $N_{R}=N_{T}=1$ and let $s = \theta \widetilde{\rho}^{-1} \widetilde{R}^{\alpha}$. The lower bound for an achievable throughput in the first hop is given by
\begin{align} \label{eq:T_FD}
\hspace{-3mm} \mathrm{T}_{\mathrm{FD}}^{(\min)}(s) \triangleq \frac{2}{(1 + s b \rho)^{a}} \exp \big( - \lambda \Upsilon^{(\max)}(s) \big) \log_{2}(1+\theta)
\end{align}
whereas, if the FD nodes operate in HD mode, the upper bound for an achievable throughput is given by
\begin{align} \label{eq:T_HD}
\mathrm{T}_{\mathrm{HD}}^{(\max)}(s) \triangleq \exp \bigg( - \lambda (1 + \tfrac{2}{\alpha}) \widetilde{\rho} \frac{\pi^{2} s^{\frac{2}{\alpha}}}{\alpha \sin \big( \frac{2 \pi}{\alpha} \big)} \bigg) \log_{2}(1+\theta).
\end{align}
Then, $\mathrm{T}_{\mathrm{FD}}^{(\min)}(s) \geq \mathrm{T}_{\mathrm{HD}}^{(\max)}(s)$ whenever the following condition holds:
\begin{align}
(1 + s b \rho)^{a} \leq 2 \exp \bigg( - \lambda \big( \big( 1 - \tfrac{2}{\alpha} \big) \widetilde{\rho} + 2 \rho \big) \frac{\pi^{2} s^{\frac{2}{\alpha}}}{\alpha \sin \big( \frac{2 \pi}{\alpha} \big)} \bigg).
\end{align}
}
\end{corollary} \vspace{-2mm}

\subsection{Second Hop} \label{sec:SP2}

In this section, we analyze the success probability of the second hop $\mathrm{P}_{\mathrm{suc}}^{(2)} = \Pr (\mathrm{SINR}_{\widehat{m}_{k}} > \theta)$, i.e., the probability of successful transmission from FD node $k$ to HD receiving node $\widehat{m}_{k}$. First, considering $\mathrm{SINR}_{\widehat{m}_{k}}$ in \eqref{eq:SINR_2}, we have $S_{k \widehat{m}_{k}} \sim \chi_{2 N_{T}}^{2}$ (desired signal) and $S_{x \widehat{m}_{k}} \sim \chi_{2}^{2}$, $\forall x \in (\Phi \backslash \{ k \}) \cup \widetilde{\Phi}$ (interferers).

The success probability of the second hop is given next in Theorem~\ref{th:P_suc2}, whereas its bounds are provided in Corollary~\ref{cor:P_suc2}.

\begin{theorem} \label{th:P_suc2} \rm{
The success probability of the second hop is given by \vspace{-1mm}
\begin{align} \label{eq:P_suc2}
\mathrm{P}_{\mathrm{suc}}^{(2)} \triangleq \sum_{n=0}^{N_{T} - 1} \bigg[ \frac{(- s)^{n}}{n!} \frac{\mathrm{d}^{n}}{\mathrm{d} s^{n}} \mathcal{L}_{I_{\widehat{m}_{k}}}(s) \bigg]_{s = \theta \rho^{-1} \widehat{R}^{\alpha}}
\end{align}
where \vspace{-2mm}
\begin{align} \label{eq:LI_2a}
\mathcal{L}_{I_{\widehat{m}_{k}}}(s) \triangleq \Psi(s, \widehat{R}) \exp \big( -\lambda \Upsilon(s) \big)
\end{align}
is the Laplace transform of $I_{\widehat{m}_{k}}$ (cf. \eqref{eq:I_2}), with $\Upsilon(s)$ and $\Psi (s, r)$ defined in \eqref{eq:Upsilon} and in\eqref{eq:Psi}, respectively.
}
\end{theorem}

\begin{IEEEproof}
See Appendix~\ref{sec:A_P_suc2_th}.
\end{IEEEproof}

\begin{corollary} \label{cor:P_suc2} \rm{
The Laplace transform of $I_{\widehat{m}_{k}}$ in \eqref{eq:LI_2a} is bounded as $\mathcal{L}_{I_{\widehat{m}_{k}}}(s) \in \big[ \mathcal{L}_{I_{\widehat{m}_{k}}}^{(\min)}(s), \mathcal{L}_{I_{\widehat{m}_{k}}}^{(\max)}(s) \big]$, with

\begin{align}
\label{eq:LI_2min} \hspace{-1mm} \mathcal{L}_{I_{\widehat{m}_{k}}}^{(\min)}(s) & \triangleq \frac{1}{1 + s \widetilde{\rho} (\widetilde{R} - \widehat{R})^{-\alpha}} \exp \big( - \lambda \Upsilon^{(\max)}(s) \big) \\
\label{eq:LI_2max} \hspace{-1mm} \mathcal{L}_{I_{\widehat{m}_{k}}}^{(\max)}(s) & \triangleq \frac{1}{1 + s \widetilde{\rho} (\widetilde{R} + \widehat{R})^{-\alpha}} \exp \big( - \lambda \Upsilon^{(\min)}(s) \big)
\end{align}
with $\Upsilon^{(\min)}(s)$ and $\Upsilon^{(\max)}(s)$ defined in \eqref{eq:Upsilon_min} and in \eqref{eq:Upsilon_max}, respectively. Then, the lower and upper bounds of the success probability of the second hop $\mathrm{P}_{\mathrm{suc}}^{(2)}$ are obtained by substituting $\mathcal{L}_{I_{\widehat{m}_{k}}}(s)$ in \eqref{eq:P_suc2} with $\mathcal{L}_{I_{\widehat{m}_{k}}}^{(\min)}(s)$ and $\mathcal{L}_{I_{\widehat{m}_{k}}}^{(\max)}(s)$, respectively.
}
\end{corollary}

\begin{IEEEproof}
See Appendix~\ref{sec:A_P_suc2_cor}.
\end{IEEEproof}

\begin{figure}[t!]
\centering
\includegraphics[scale=1]{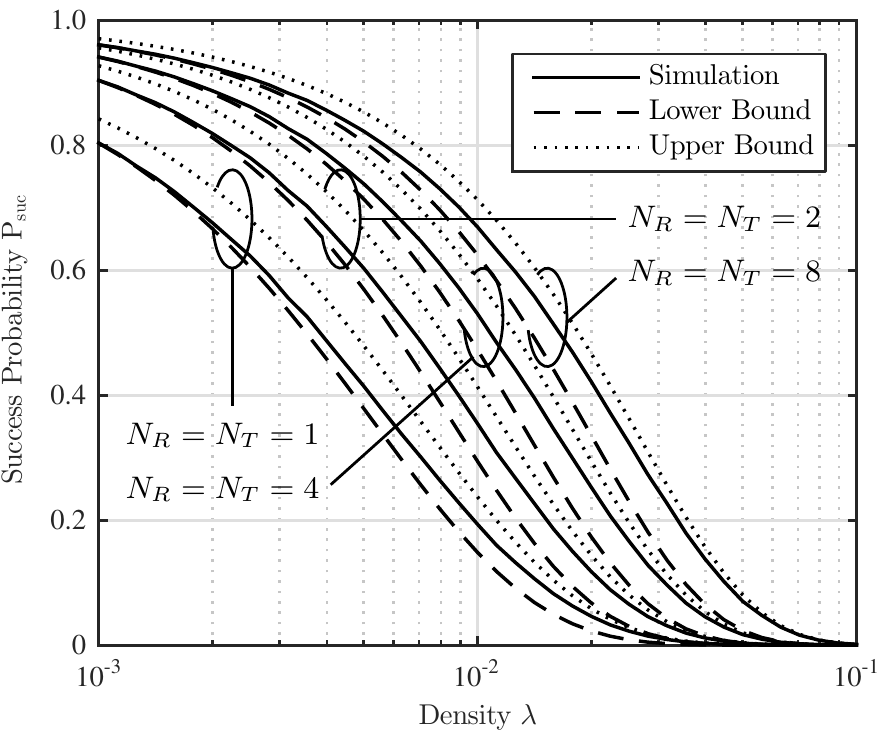}
\caption{Success probability: simulations and theoretical bounds for different values of the spatial density $\lambda$, with $\Omega=-80$~dB and $\theta=0$~dB.} \label{fig:P_suc} \vspace{-10mm}
\end{figure}

\section{Numerical Results} \label{sec:Num}

In this section, we present numerical results to assess our theoretical findings and, specifically, to compare the performance of FD with respect to HD. In the following, the pathloss exponent is $\alpha=4$, the distances characterizing the marked PPP are set to $\widetilde{R}=5$~m and $\widehat{R}=0.5$~m, the transmit powers are $\rho=0.5$~W and $\widetilde{\rho}=1$~W, and the considered SINR threshold is $\theta=0$~dB; the parameters $a$ and $b$ of the self-interference power are computed according to \eqref{eq:ab}, where $\mu$ and $\nu$ are obtained from \eqref{eq:mu_nu} with Ricean $K$-factor $K=1$ (see \cite{Dua12} for an experimental characterization of $K$) and self-interference attenuation $\Omega=-80$~dB. Lastly, the FD nodes adopt maximum ratio combining and maximum ratio transmission and, therefore, the beamforming vectors are given by \vspace{-2mm}
\begin{align} \label{eq:MRC_MRT}
\v_{z} = \frac{\h_{\widetilde{m}_{z} z}}{\| \h_{\widetilde{m}_{z} z} \|}, \qquad \w_{x} = \frac{\h_{x \widehat{m}_{x}}}{\| \h_{x \widehat{m}_{x}} \|}.
\end{align}

Figure~\ref{fig:P_suc} plots the success probability $\mathrm{P}_{\mathrm{suc}}$ in \eqref{eq:P_suc} over different values of the density $\lambda$. Remarkably, the theoretical bounds obtained in the previous section are reasonably tight. Furthermore, the employment of multiple antennas allows to mitigate the effect of the self-interference of the FD nodes, thus producing substantial SINR gains (observe that to higher values of $N_{R}$ correspond more terms in the summation \eqref{eq:P_suc1}).

\begin{figure}[t!]
\centering
\includegraphics[scale=1]{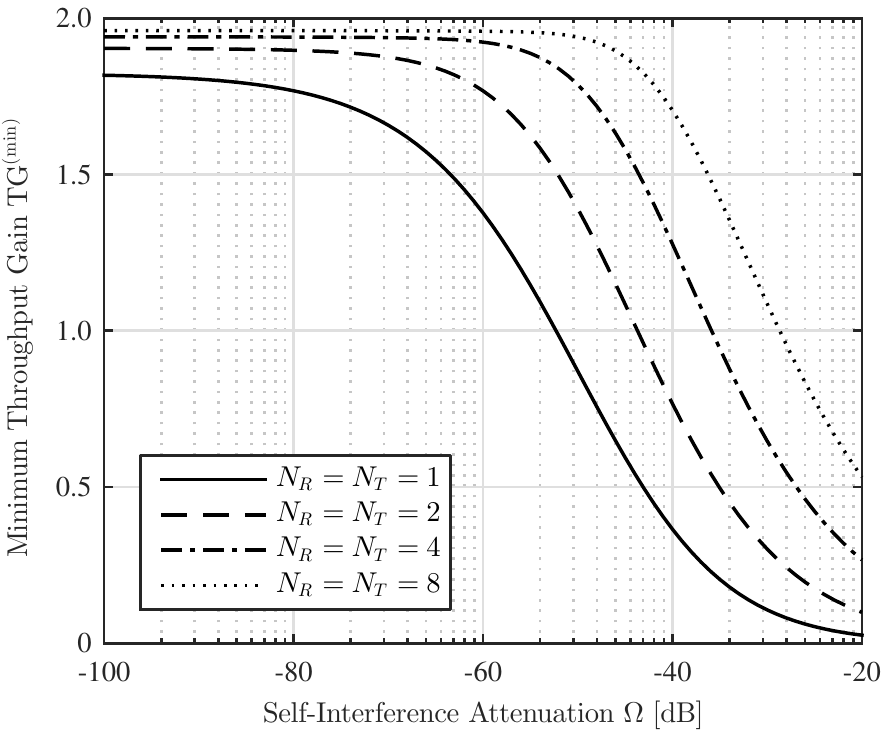}
\caption{Comparison between FD and HD modes in the first hop: minimum throughput gain for different values of the self-interference attenuation $\Omega$, with $\lambda=10^{-3}$ and $\theta=0$~dB.} \label{fig:TG_Omega} \vspace{-3mm}
\end{figure}

\begin{figure}[t!]
\centering
\includegraphics[scale=1]{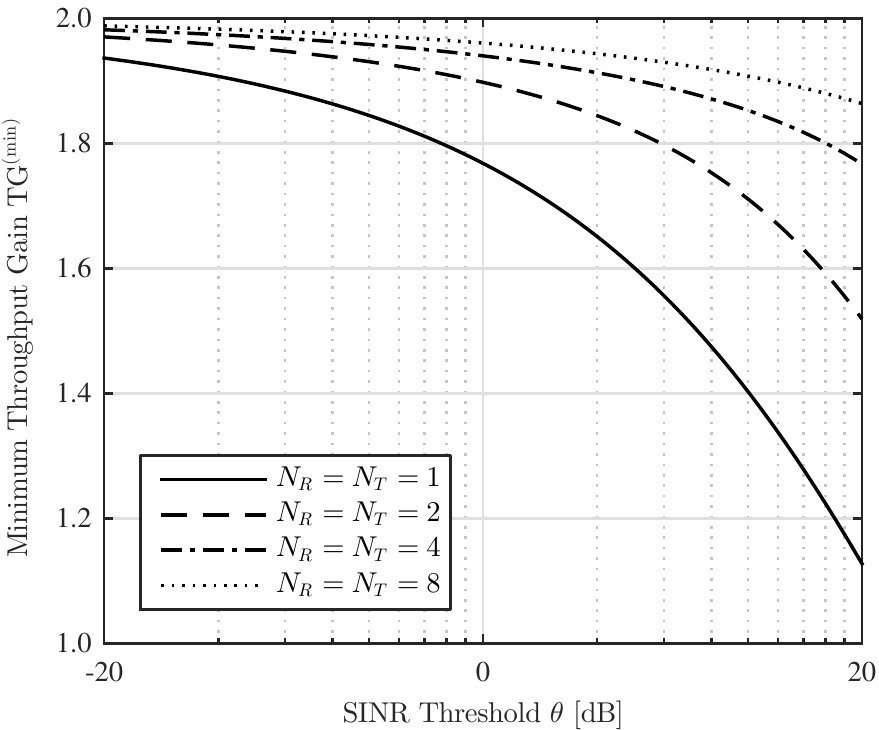}
\caption{Comparison between FD and HD modes in the first hop: minimum throughput gain for different values of the SINR threshold $\theta$, with $\lambda=10^{-3}$ and $\Omega=-80$~dB.} \label{fig:TG_theta} \vspace{-4mm}
\end{figure}

We now focus our attention on the first hop (subject to self-interference) in order to analyze the benefits of FD. With this objective in mind, we introduce the minimum throughput gain
\begin{align}
\mathrm{TG}^{(\min)} \triangleq \frac{\mathrm{T}_{\mathrm{FD}}^{(\min)}}{\mathrm{T}_{\mathrm{HD}}^{(\max)}},
\end{align}
with $\mathrm{T}_{\mathrm{FD}}^{(\min)}$ and $\mathrm{T}_{\mathrm{HD}}^{(\max)}$ defined in \eqref{eq:T_FD} and \eqref{eq:T_HD}, respectively: this parameter denotes the worst-case gain of FD mode over HD mode in terms of throughput, with $\mathrm{TG}^{(\min)} > 1$ indicating that FD outperforms the equivalent HD setup. Figure~\ref{fig:TG_Omega} plots $\mathrm{TG}^{(\min)}$ over different values of the self-interference attenuation $\Omega$ with $\lambda=10^{-3}$, whereas all the other parameters are the same as in the previous simulation. In this setting, we have $\mathrm{TG}^{(\min)} \geq 1$ even for moderate values of the attenuation and, in the specific, when: $\Omega \leq -52$~dB for $N_{R}=N_{T}=1$, $\Omega \leq -44$~dB for $N_{R}=N_{T}=2$, $\Omega \leq -36$~dB for $N_{R}=N_{T}=4$, and $\Omega \leq -28$~dB for $N_{R}=N_{T}=8$. Moreover, the minimum throughput gain is analyzed in Figure~\ref{fig:TG_theta} as a function of the SINR threshold $\theta$ with $\lambda=10^{-3}$ and $\Omega=-80$~dB: in this respect, it is shown that FD achieves improved performance with respect to HD for any reasonable value of $\theta$.

\section{Conclusions} \label{sec:Concl}

This paper analyzes the performance of wireless networks with FD MIMO small cells using stochastic geometry. We characterize the distribution of the self-interference power of the FD nodes for arbitrary receive/transmit beamforming strategies and we derive tight bounds for the success probability. Our framework highlights the beneficial effect of FD, which produces substantial throughput gains over HD for realistic values of number of antennas, node density, self-interference attenuation, and SINR threshold.

\appendices

\section{Success Probability of the First Hop}
\subsection{Proof of Lemma~\rm{\ref{lem:SI}}} \label{sec:A_SI}

Due to space limitations, we only provide here a sketch of the proof. For notational simplicity, in the following we omit the sub-indices in the beamforming vectors and in the channel matrix and write $S_{k k} \triangleq |\v^{\herm} \H \w|^{2}$, with
\begin{align} \label{eq:vHw}
|\v^{\herm} \H \w|^{2} = \sum_{i,k=1}^{N_{R}} \sum_{j,\ell=1}^{N_{T}} v_{i}^{*} v_{k} h_{i j} h_{k \ell}^{*} w_{j} w_{\ell}^{*}
\end{align}
where $v_{i}$ is the $i$-th element of $\v \triangleq (v_{i})_{i=1}^{N_{R}}$, $w_{j}$ is the $j$-th element of $\w \triangleq (w_{j})_{j=1}^{N_{T}}$, and $h_{i j}$ is the $(i,j)$-th element of $\H \triangleq \big( (h_{i j})_{i=1}^{N_{R}} \big)_{j=1}^{N_{T}}$. Recalling that $h_{i j} \sim \setC \setN (\mu, \nu^{2})$, we have $\Exp \big( |h_{i j}|^{2} \big) = \mu^{2} + \nu^{2}$ and $\Exp \big( |h_{i j}|^{4} \big) = \mu^{4} + 4 \mu^{2} \nu^{2} + 2 \nu^{4}$; on the other hand, for any normalized receive/transmit beamforming vector, we have $\Exp \big( |v_{i}|^{4} \big) = \frac{1}{N_{R} (N_{R} + 1)}$ and $\Exp \big( |w_{j}|^{4} \big) = \frac{1}{N_{T} (N_{T} + 1)}$. Building on the central limit theorem for causal functions \cite{Pap62}, we approximate a sum of positive random variables $x = \sum_{i} x_{i}$ by a Gamma distribution with shape parameter $a$ and scale parameter $b$ defined, respectively, as
\begin{align} \label{eq:gamma_ab}
a \triangleq \frac{(\Exp(x))^{2}}{\Var(x)}, \qquad a \triangleq \frac{\Var(x)}{\Exp(x)}.
\end{align}

\begin{figure*}[t!]
\newcounter{eq2}
\setcounter{eq2}{\value{equation}}
\begin{align}
\addtocounter{equation}{+6}
\label{eq:LI_1b} \mathcal{L}_{I_{k}}(s) & = \Exp \big( \exp (- s \rho S_{k k}) \big) \Exp \bigg( \prod_{\ell \in \Phi \backslash \{k\}} \exp \big( - s (\rho |X_{\ell}|^{-\alpha} S_{\ell k} + \widetilde{\rho} |X_{\widetilde{m}_{\ell}}|^{-\alpha} S_{\widetilde{m}_{\ell} k}) \big) \bigg) \\
\addtocounter{equation}{+2}
\label{eq:LI_2b} \mathcal{L}_{I_{\widehat{m}_{k}}}(s) & = \Exp \big( \exp (- s \widetilde{\rho} |X_{\widetilde{m}_{k}}|^{-\alpha} S_{\widetilde{m}_{k} \widehat{m}_{k}}) \big) \Exp \bigg( \prod_{l \in \Phi \backslash \{k\}} \exp \big( - s (\rho |X_{\ell}|^{-\alpha} S_{\ell \widehat{m}_{k}} + \widetilde{\rho} |X_{\widetilde{m}_{\ell}}|^{-\alpha} S_{\widetilde{m}_{\ell} \widehat{m}_{k}}) \big) \bigg)
\end{align}
\setcounter{equation}{\value{eq2}}
\hrulefill
\vspace{-6mm}
\end{figure*}

In order to compute $a$ and $b$, we need to derive the second and fourth moments of $|\v^{\herm} \H \w|^{2}$. Let $\sigma_{i}$ denote the $i$-th singular value of $\H$ and $N_{\min} \triangleq \min(N_{R}, N_{T})$: from \eqref{eq:vHw}, after some algebraic manipulations we obtain
\begin{align}
\hspace{-2mm} \Exp \big( |\v^{\herm} \H \w|^{2} \big) & = \Exp \bigg( \sum_{i=1}^{N_{R}} \sum_{j=1}^{N_{T}} |v_{i}|^{2} |h_{i j}|^{2} |w_{j}|^{2} \bigg) \\ 
& = \mu^{2} + \nu^{2}
\end{align}
and \vspace{-3mm}
\begin{align}
\hspace{-2mm} \nonumber \Exp \big( |\v^{\herm} \H \w|^{4} \big) & = \Exp \bigg( \sum_{i=1}^{N_{\min}} |v_{i}|^{4} |\sigma_{i}|^{4} |w_{j}|^{4} \bigg) \\
\hspace{-2mm} & \hspace{-5mm} + 2 \Exp \bigg( \sum_{\substack{i,j=1 \\ i \neq j}}^{N_{\min}} |v_{i}|^{2} |v_{j}|^{2} |\sigma_{i}|^{2} |\sigma_{j}|^{2} |w_{i}|^{2} |w_{j}|^{2} \bigg) \\
\hspace{-2mm} & = \frac{4 N_{R} N_{T}}{(N_{R} + 1) (N_{T} + 1)} \mu^{4} + 4 \mu^{2} \nu^{2} + 2 \nu^{4}.
\end{align}
Since $\Var \big( |\v^{\herm} \H \w|^{2} \big) = \Exp \big( |\v^{\herm} \H \w|^{4} \big) - \big( \Exp \big( |\v^{\herm} \H \w|^{2} \big) \big)^{2}$, we readily obtain $a$ and $b$ in \eqref{eq:ab} from \eqref{eq:gamma_ab}. \hfill \IEEEQED

\subsection{Proof of Theorem~\rm{\ref{th:P_suc1}}} \label{sec:A_P_suc1_th}

The success probability of the first hop is given by
\begin{align}
\mathrm{P}_{\mathrm{suc}}^{(1)} & = \Pr \bigg( \frac{\widetilde{\rho} \widetilde{R}^{-\alpha} S_{\widetilde{m}_{k} k}}{I_{k}} > \theta \bigg) \\
& = \Pr \big( S_{\widetilde{m}_{k} k} > \theta \widetilde{\rho}^{-1} \widetilde{R}^{\alpha} I_{k} \big)
\end{align}
where $I_{k}$ is defined in \eqref{eq:I_1} and denotes the overall interference at $k$. Since FD node $k$ is equipped with $N_{R}$ receive antennas, the power of its desired signal is distributed as $S_{\widetilde{m}_{k} k} \sim \chi_{2 N_{R}}^{2}$ and our case falls into the general framework \cite{Hun08}; hence, the expression in \eqref{eq:P_suc1} results from applying \cite[Th.~1]{Hun08}. On the other hand, the Laplace transform of $I_{k}$ is obtained as in \eqref{eq:LI_1b} at the top of the page using the moment-generating function of the Gamma and of the $\chi_{2}^{2}$ distributions. Finally, applying \cite[Th.~1]{Ton15}, the expression in \eqref{eq:LI_1a} readily follows. \hfill \IEEEQED

\subsection{Proof of Corollary~\rm{\ref{cor:P_suc1}}} \label{sec:A_P_suc1_cor}

Building on \cite[Th.~2]{Ton15}, we obtain that $\Upsilon(s)$ in \eqref{eq:Upsilon} is bounded as $\Upsilon(s) \in \big[ \Upsilon^{(\min)}(s), \Upsilon^{(\max)}(s) \big]$, with $\Upsilon^{(\max)}(s)$ and $\Upsilon^{(\min)}(s)$ defined in \eqref{eq:Upsilon_min} and in \eqref{eq:Upsilon_max}, respectively, from which we immediately derive the lower and upper bounds of $\mathcal{L}_{I_{k}}(s)$ in \eqref{eq:LI_1min}--\eqref{eq:LI_1max}. \hfill \IEEEQED \vspace{-1mm}

\section{Success Probability of the Second Hop}
\subsection{Proof of Theorem~\rm{\ref{th:P_suc2}}} \label{sec:A_P_suc2_th}

The success probability of the second hop is given by
\addtocounter{equation}{+1}
\begin{align}
\mathrm{P}_{\mathrm{suc}}^{(2)} & = \Pr \bigg( \frac{\rho \widehat{R}^{-\alpha} S_{k \widehat{m}_{k}}}{I_{\widehat{m}_{k}}} > \theta \bigg) \\
& = \Pr \big( S_{k \widehat{m}_{k}} > \theta \rho^{-1} \widehat{R}^{\alpha} I_{\widehat{m}_{k}} \big)
\end{align}
where $I_{\widehat{m}_{k}}$ is defined in \eqref{eq:I_2} and denotes the overall interference at $\widehat{m}_{k}$. Since FD node $k$ is equipped with $N_{T}$ transmit antennas, the power of the desired signal of $\widehat{m}_{k}$ is distributed as $S_{k \widehat{m}_{k}} \sim \chi_{2 N_{T}}^{2}$ and, similarly as in Appendix~\ref{sec:A_P_suc1_th}, the expression in \eqref{eq:P_suc2} results from applying \cite[Th.~1]{Hun08}. On the other hand, the Laplace transform of $I_{\widehat{m}_{k}}$ is obtained as in \eqref{eq:LI_2b} at the top of the page using the moment-generating function of the $\chi_{2}^{2}$ distribution. Again, we resort to \cite[Th.~2]{Ton15} and the expression in \eqref{eq:LI_2a} readily follows. \hfill \IEEEQED

\subsection{Proof of Corollary~\rm{\ref{cor:P_suc2}}} \label{sec:A_P_suc2_cor}

Given the definition of $\Psi(s,r)$ in \eqref{eq:Psi}, we observe that
\addtocounter{equation}{+1}
\begin{align} \label{eq:Psi_bounds}
\Psi(s,\widehat{R}) \in \bigg[ \frac{1}{1 + s \widetilde{\rho} (\widetilde{R} - \widehat{R})^{-\alpha}} , \frac{1}{1 + s \widetilde{\rho} (\widetilde{R} + \widehat{R})^{-\alpha}} \bigg]
\end{align}
where we recall that $\widetilde{R} \gg \widehat{R}$. Then, the lower and upper bounds of $\mathcal{L}_{I_{\widehat{m}_{k}}}(s)$ in \eqref{eq:LI_2min}--\eqref{eq:LI_2max} are a straightforward result of combining \eqref{eq:Psi_bounds} and Corollary~\ref{cor:P_suc1}. \hfill \IEEEQED

%

\addcontentsline{toc}{chapter}{References}
\bibliographystyle{IEEEtran}
\bibliography{IEEEabrv,ref_Huawei}

\end{document}

%% file: Definitions.tex

\newcommand{\h}{\mathbf{h}}

\newcommand{\n}{\mathbf{n}}

\renewcommand{\v}{\mathbf{v}}
\newcommand{\w}{\mathbf{w}}

\newcommand{\y}{\mathbf{y}}


\renewcommand{\H}{\mathbf{H}}




\newcommand{\setC}{\mathcal{C}}

\newcommand{\setN}{\mathcal{N}}

\newcommand{\Real}{\mbox{$\mathbb{R}$}}
\newcommand{\Compl}{\mbox{$\mathbb{C}$}}


\newcommand{\Exp}{\mathbb{E}}
\newcommand{\herm}{\mathrm{H}}
\renewcommand{\Pr}{\mathbb{P}}

\newcommand{\Var}{\mathrm{Var}}